\documentclass[final,5p,times,twocolumn,authoryear]{elsarticle}

\usepackage[authoryear]{natbib}
\usepackage{graphicx}
\usepackage{subcaption}
\usepackage{amssymb,amsmath}
\usepackage{pdflscape}
\usepackage{adjustbox}

\usepackage{booktabs}
\usepackage{lipsum}
\usepackage{epsfig}
\usepackage{txfonts}
\usepackage{xcolor}
\usepackage{tabularx}
\usepackage{footnote}
\usepackage{soul}
\usepackage{multirow}
\usepackage{multicol}
\usepackage{mathpazo}
\usepackage{lineno,hyperref}
\usepackage{textcomp}
\usepackage[utf8]{inputenc}
\usepackage[T1]{fontenc}
\usepackage{textgreek} 
\usepackage[utf8]{inputenc}
\usepackage{booktabs}

\usepackage{amsthm}

\journal{High Energy Astrophysics}

\begin{document}

\begin{frontmatter}



\title{A Possible Short-Timescale Optical Quasi-Periodic Oscillation in PKS\,0805$-$07
from High-Cadence TESS Observations}


%

\author{Sikandar Akbar\corref{cor1}\fnref{label1}}
\ead{darprince46@gmail.com}

\author{Zahir Shah\corref{cor2}\fnref{label2}}
\ead{shahzahir4@gmail.com}

\author{Athar A. Dar\corref{cor3}\fnref{label3}}
\ead{atherdar6@gmail.com}

\author{Naseer Iqbal\fnref{label1}}

\affiliation[label1]{Department of Physics, University of Kashmir,
Srinagar 190006, India}

\affiliation[label2]{Manipal Centre for Natural Sciences, Centre of
Excellence, Manipal Academy of Higher Education, Manipal 576104, India}

\affiliation[label3]{Department of Physics, Central University of
Kashmir, Ganderbal 191201, India}

\cortext[cor1]{Corresponding author: Sikandar Akbar.
Email: darprince46@gmail.com}

\cortext[cor2]{Corresponding author: Zahir Shah.
Email: shahzahir4@gmail.com}

\cortext[cor3]{Corresponding author: Athar A. Dar.
Email: atherdar6@gmail.com}

\begin{abstract}

We present a timing analysis of the high-cadence optical 
light curve of the high-redshift flat-spectrum radio quasar 
PKS\,0805$-$07 obtained during \textit{TESS} Sector~34 
(MJD~=~59230.90--59239.90). We search for short-timescale 
quasi-periodic oscillations (QPOs) using complementary 
time-series techniques, including the Lomb--Scargle 
periodogram (LSP) and the weighted wavelet $Z$-transform 
(WWZ), and evaluate their significance against red-noise 
variability using Monte Carlo simulations. The LSP reveals 
a dominant modulation at $f \approx 0.6\,\mathrm{d^{-1}}$ 
($P \approx 1.7$\,d) exceeding the $99.99\%$ confidence 
level, while the WWZ independently recovers a consistent 
timescale at the $\sim$99.9\% level and shows that the 
signal is temporally localized rather than persistent across 
the full light curve. The modulation spans $\sim$5 coherent 
cycles, indicating a transient quasi-periodic feature. 
However, the detected significance only marginally exceeds 
the highest confidence threshold in the WWZ analysis, and 
the limited number of coherent cycles is insufficient for 
a firm confirmation.
We discuss possible physical interpretations of the 
candidate modulation. In a disk-based scenario, orbital 
motion of a hotspot near the innermost stable circular 
orbit implies a black hole mass of 
$M_{\rm BH} \sim 7.2 \times 10^{8}\,M_\odot$, consistent 
with typical FSRQ values. Alternatively, magnetohydrodynamic 
kink instabilities in the relativistic jet can naturally 
produce day-scale variability for standard blazar parameters 
and account for the transient character of the signal. 
Compact SMBH binary and tidal disruption event scenarios 
are also considered but found to be disfavoured by the 
jet-dominated nature of the source, the optical rather than 
X-ray character of the modulation, and the black hole mass 
well above the regime associated with confirmed 
QPE host galaxies. We conclude that the candidate 
modulation is consistent with a compact, short-lived 
structure embedded within stochastic jet variability, 
and that high-cadence multiwavelength monitoring will be 
essential to confirm its recurrence and constrain its 
physical origin.

\end{abstract}



\end{frontmatter}




\section{Introduction}
\label{introduction}

PKS\,0805$-$07 is a high-redshift ($z=1.837$; \citealt{1988ApJ...327..561W})
flat-spectrum radio quasar (FSRQ) known for its pronounced $\gamma$-ray
output and repeated flaring episodes \citep{2024ApJ...977..111A}. FSRQs
are part of the blazar class of active galactic nuclei (AGN), distinguished
by relativistic jets oriented nearly along our line of sight. This alignment
results in strong Doppler boosting and pronounced flux variability, making
such sources excellent probes for studying extreme variability and
jet-related physical mechanisms.

Quasi-periodic oscillations (QPOs) have been reported in AGNs across the
entire electromagnetic spectrum, from radio to $\gamma$-rays, with claimed
timescales ranging from minutes to decades. On minute-to-hour scales, early
detections include periodicities in OJ~287
\citep{1985Natur.314..148V,1985Natur.314..146C}. On day scales, QPOs have
been identified in several AGNs using space-based high-cadence photometry,
including modulations in CTA~102, PKS~1510$-$089, and BL~Lacertae
\citep{Sarkar2020,Roy2022,2022Natur.609..265J,2024MNRAS.527.9132T}.
On month scales, periodicities of $\sim$30--120\,days have been reported
in PKS~2247$-$131 \citep{Zhou2018}, B2~1520+31 \citep{Gupta2019}, and
J1359+4011 \citep{King2013}. On year-like timescales, QPOs have been
found in AO~0235+16 \citep{Raiteri2001}, 3C~66A \citep{OteroSantos2020},
PKS~J2134$-$0153 \citep{Ren2021}, and PKS~0405$-$385 \citep{Gong2022}.
In the $\gamma$-ray band, the first QPO was reported in PG~1553+113 at
$\sim$2\,yr \citep{Ackermann2015}, and more than 30 QPO signals have since
been identified in blazar $\gamma$-ray light curves with periods from months
to years \citep{Sandrinelli2016,2017MNRAS.471.3036P,Zhang2017,Bhatta2019,
Zhang2021,Zhang2023}.

Among the physical mechanisms proposed to explain QPOs, the SMBHB scenario
is among the most widely discussed, in which the periodicity reflects the
Keplerian orbital period of a secondary black hole \citep{Komossa2006}.
The most established example is OJ~287, exhibiting a $\sim$12\,yr optical
QPO attributed to a binary system with separation $\sim$0.1\,pc
\citep{Sillanpaa1988,Valtonen2006}; year-like periodicities in other
blazars are similarly attributed to SMBHB systems, helical jet motion,
or jet precession
\citep[e.g.,][]{1996A&A...305L..17S,2015Natur.518...74G,
2022MNRAS.513.5238R,AKBAR2026100608}.
SMBHB-induced QPOs are expected on timescales of $\sim$1--25\,yr
\citep{Komossa2006} and provide an important channel for identifying
gravitational wave sources detectable by Pulsar Timing Arrays
\citep{Agazie2023}. On shorter timescales, additional mechanisms
include the helical jet model \citep{Camenzind1992,Mohan2015},
Lense--Thirring precession of a misaligned accretion disc \citep{Liska2018},
accretion disc instabilities \citep{BhattaDhital2020}, diskoseismic modes,
and magnetic reconnection within the jet \citep{Huang2013}. Since
characteristic disc oscillation frequencies scale inversely with black
hole mass, QPOs also provide a potential probe of accretion physics
across a wide mass range and an indirect method for mass estimation.

Detecting optical QPOs in ground-based data is challenging because uneven sampling, seasonal gaps, and red-noise variability can mimic or obscure periodic signals \citep{2016MNRAS.461.3145V}. Space-based missions such as \textit{Kepler} \citep{2010Sci...327..977B} and the Transiting Exoplanet Survey Satellite (TESS,\citealt{2015JATIS...1a4003R}), have significantly improved this situation by delivering high-precision, uniformly sampled, high-cadence light curves that enable more reliable periodicity searches. Such datasets make it possible to probe short-to-intermediate timescales with reduced systematic biases and enhanced statistical confidence. Proposed physical mechanisms for optical and multiwavelength QPOs include disk oscillations, warped or precessing accretion flows, disk--jet coupling, and magnetically driven processes such as reconnection-induced flux-rope formation in relativistic jets \citep[e.g.,][]{1999ApJ...524L..63S,2001ApJ...559L..25W,2005PASJ...57..699K,2022ApJ...933...55C}. Despite these advances, the origin of QPOs in AGNs remains an open problem.

PKS\,0805$-$07 has recently emerged as a promising candidate for quasi-periodic variability in blazars. In our previous study \citep{2025PhRvD.112f3061A}, we carried out a multi-technique time-series analysis of its long-term Fermi-LAT $\gamma$-ray light curve and reported indication for dual quasi-periodic oscillations at $\sim255$ and $\sim112$\,days during the active interval MJD~59047.5--59740.5. These signals were detected with multiple independent techniques and supported by phase-folding and model-selection tests, indicating amplitude-modulated variability plausibly linked to geometric effects such as jet precession coupled with a secondary dynamical process. While these results provided indication for complex periodic behavior
in the $\gamma$-ray band, an important open question is whether similar
signatures are present on different timescales and in other wavebands.

In this work, we extend the investigation to high-cadence optical observations from TESS in order to probe day-scale variability in PKS\,0805$-$07 and examine the multiwavelength manifestation of the candidate periodic processes. Establishing (or refuting) a correspondence between optical and $\gamma$-ray variability is essential for distinguishing between geometric scenarios---such as Doppler-factor modulation from jet precession---and intrinsic radiative or particle-acceleration mechanisms, since the former can produce correlated modulation across bands while the latter is expected to be more energy dependent. 
To this end, we apply complementary time-series techniques, including
the Lomb--Scargle periodogram \citep[LSP;][]{lomb1976least, scargle1982studies},
the weighted wavelet Z-transform \citep[WWZ;][]{foster1996wavelets}, and epoch
folding \citep{Leahy1983, Larsson1996}, allowing us to investigate both
the frequency content and the temporal localization of the modulation
while accounting for red-noise variability and the finite duration of a
single TESS sector.

The paper is organized as follows. Section~\ref{sec:1} describes the
data selection and reduction procedures. Section~\ref{QPO} presents
the timing-analysis results for the TESS light curve.
Section~\ref{sig_ev} outlines the statistical significance tests for
the LSP and WWZ analyses. Section~\ref{sec:sum} summarizes the main
findings and discusses their physical implications.

\begin{figure*}
\centering
\includegraphics[width=0.9\textwidth]{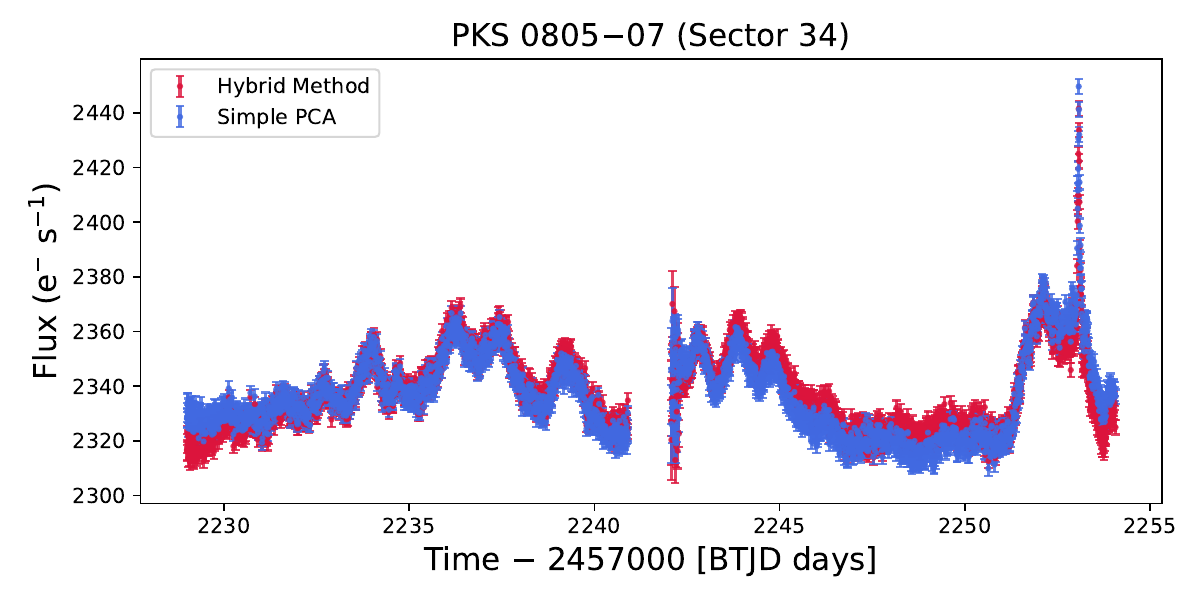}
\caption{Systematics-corrected TESS Sector~34 optical light curve of PKS\,0805$-$07 obtained using the \textsc{Quaver} pipeline. The red points correspond to the hybrid reduction, while the blue points show the Simple PCA (SPO) light curve adopted for the timing analysis. The time axis is given in BTJD ($\mathrm{BTJD} = \mathrm{BJD} - 2457000$), and the flux is in instrumental $\mathrm{e^{-}\,s^{-1}}$. }
\label{fig:sector34_lc}

\end{figure*}

\section{Observations and Data Reduction}
\label{sec:1}

\subsection{TESS Observations}
\label{sec:obs}

The optical data used in this work were obtained with the \textit{Transiting Exoplanet Survey Satellite} (TESS; \citealt{2015JATIS...1a4003R}), a space-based mission designed for high-precision, time-domain photometry over nearly the full sky. TESS carries four wide-field CCD cameras, each covering $24^{\circ}\times24^{\circ}$, providing a combined instantaneous field of view of $24^{\circ}\times96^{\circ}$. The survey strategy divides the sky into 26 ``sectors'' (13 per hemisphere), with each sector monitored almost continuously for $\sim$27~days. The observing cadence is 30\,min in the early mission (2018--2019) and 10\,min or 2\,min in later cycles (2020--present), enabling detailed studies of short-timescale variability. Unlike ground-based optical monitoring, TESS light curves are free from day--night interruptions and seasonal gaps, resulting in nearly continuous coverage and significantly improved sampling. The effective continuous baseline for a given source depends on its ecliptic latitude, ranging from a single-sector duration near the ecliptic plane to almost one year in the continuous viewing zones near the ecliptic poles. After each year-long observing cycle the spacecraft swaps hemispheres and repeats the survey, so that some sources are observed in multiple sectors separated by approximately one year. The photometric precision of TESS is magnitude dependent; for sources of comparable brightness to PKS\,0805$-$07, variability at the $\sim1$--10\% level can be reliably detected.

PKS\,0805$-$07 was observed by TESS in Sector~34 (Cycle~3), covering the interval MJD~59229--59254 (14~January~2021 to 8~February~2021) with a 10\,min cadence. The source has a TESS magnitude of
$T \approx 17.7$\,mag ($V \approx 18.5$\,mag) and was observed in
the TESS broad red-optical bandpass spanning $600$--$1000$\,nm,
centred on the Cousins $I$-band at a central wavelength of
$786.5$\,nm \citep{2015JATIS...1a4003R}. The target coordinates are $\alpha_{\rm J2000}=122.06473349^{\circ}$ and $\delta_{\rm J2000}=-7.85274626^{\circ}$, as returned by our extraction query. In addition to Sector~34, the source is also available in other TESS
sectors (e.g., Sectors 7, 61, 88, and 99). We examined the light curves from these additional sectors and find that they are either dominated by flaring activity, show higher noise levels, or lack the quasi-sinusoidal structure identified in Sector~34. The quasi-periodic feature therefore appears to be a transient phenomenon confined to the Sector~34 epoch, consistent
with the temporally localized nature of the signal established by the WWZ analysis (Section~\ref{QPO}). Accordingly, the present analysis is restricted to Sector~34.

\subsection{Data Reduction}
\label{sec:data_red}

We extracted and calibrated the TESS Sector~34 light curve using the open-source \textsc{Quaver} pipeline\footnote{\url{https://github.com/kristalynnesmith/quaver}} \citep{Smith_2023}, which is designed to produce systematics-corrected light curves from TESS full-frame images (FFIs) in a manner that is optimized for variability studies. \textsc{Quaver} interfaces with \textit{TESSCut} \citep{Brasseur2019}, a service provided by the Mikulski Archive for Space Telescopes (MAST)
that generates time-series cutouts of the TESS Full Frame Images (FFIs)
for any sky position or object name. Given a target coordinate,
\textit{TESSCut} extracts a compact ``postage-stamp'' sub-image centred
on the source and returns it as a target pixel file compatible with
standard TESS pipeline products, thereby avoiding the need to download
the full and substantially larger FFI data products. This cutout-based
approach is particularly advantageous in crowded fields, as it allows
the extraction aperture to be customized around the source and helps
reduce contamination from nearby objects.

A key feature of \textsc{Quaver} is its principal component analysis (PCA) framework for mitigating instrumental effects and contamination. In particular, \textsc{Quaver} applies PCA to model and subtract spacecraft systematics, variability from contaminating sources within the cutout, and scattered background light. In contrast to a purely ``simple PCA'' approach, \textsc{Quaver} adopts a hybrid philosophy in which faint background pixels can be used to track additive systematics (e.g., scattered light), while brighter pixels--- which may contain astrophysical contamination and multiplicative trends---are handled separately through the regression design matrix. Prior to applying the correction, \textsc{Quaver} prompts an interactive aperture selection step, guided by an overlay of DSS contours on the cutout, so that the user can choose an extraction aperture that minimizes blending and background leakage. The pipeline also allows optional masking of cadences affected by strong artifacts (e.g., thermal ramps or spike-like features), improving the stability of the final corrected light curve.
For the systematic correction, \textsc{Quaver} provides multiple
reduction modes, including (i) a full hybrid overfit (FHO), (ii) a
simple hybrid overfit (SHO), and (iii) a simple PCA overfit (SPO);
full details of each mode are described in \citet{Smith_2023}. Since our primary goal is timing analysis and QPO searches---where preserving intrinsic short-timescale variability while suppressing instrumental structure is essential---we adopted the PCA-based option (SPO) to generate the corrected Sector~34 light curve used throughout this work. The reduction was performed by querying the target common name
(``pks 0805-07'') and selecting Cycle~3, which returns Sector~34
among the available products. The resulting output includes the
corrected light curve and diagnostic plots (aperture selection,
regression components, and correction performance), which were
inspected to ensure that the dominant instrumental signatures were
removed without introducing spurious periodic features. The
systematics-corrected Sector~34 light curve is shown in
Figure~\ref{fig:sector34_lc}.
To verify that the detected variability is not an artifact of the
detrending procedure, we compared the SPO reduction with the hybrid
overfit products. Both reductions recover consistent variability
patterns, demonstrating that the PCA-based correction effectively
removes long-term instrumental trends while preserving the
short-timescale signal relevant for the QPO search. The final
calibrated Sector~34 time series was then used as input for the
period-search analyses presented in Section~\ref{QPO} (LSP, WWZ,
and epoch folding). The processed TESS light curve contains a short
gap of approximately one day, which is likely associated with
routine spacecraft operations such as data downlink or command
uploads. To facilitate efficient time-series analysis and reduce the
impact of high-frequency noise, we rebinned the light curve by
averaging the data points within 1.5-hour intervals and treating
each bin as a single measurement. This binning preserves the
day-scale variability of interest while reducing the data volume and
improving the computational efficiency and numerical stability of
the LSP, WWZ, and epoch folding analyses. The LSP and WWZ analyses
were carried out over the interval
$\mathrm{BTJD}=2231.4047$--$2240.4047$
($\mathrm{MJD}=59230.90$--$59239.90$), selected from the
1.5\,hr binned light curve for the QPO investigation. The epoch
folding analysis was performed over the same interval using the
same binned light curve.

\section{Quasi-periodic oscillation}
\label{QPO}
To search for quasi-periodic signatures in the  TESS light curve of PKS\,0805$-$07, we applied  LSP, WWZ and epoch folding. In addition, extensive Monte Carlo simulations were carried out to evaluate the statistical significance of the identified features. The methodologies and their associated results are described in detail in the sections that follow.

\subsection{Lomb-Scargle Periodogram (LSP)}\label{sec:lsp}

The Lomb--Scargle periodogram (LSP) is a widely used technique for identifying periodic signals in unevenly sampled time-series data \citep{lomb1976least,scargle1982studies}. Owing to its suitability for irregularly spaced light curves, it has become a standard tool in astronomical variability analyses. In this work, we employed the \textsc{Astropy} implementation of the Lomb--Scargle algorithm\footnote{\url{https://docs.astropy.org/en/stable/timeseries/lombscargle.html}}, incorporating the measured flux uncertainties to improve the stability of the resulting periodograms. Our application of the LSP follows the approach adopted in our earlier variability studies \citep{Nazir_2026,AKBAR2026100608,2025PhRvD.112f3061A}. The frequency search was restricted to the range $f_{\rm min}=1/T$ to $f_{\rm max}=1/(2\Delta T)$, where $T$ denotes the total temporal baseline of the TESS observation and $\Delta T$ the characteristic sampling interval; further details of the underlying formalism are given by \citet{vanderplas2018understanding}.

\begin{figure*}
\centering
\includegraphics[width=0.9\textwidth]{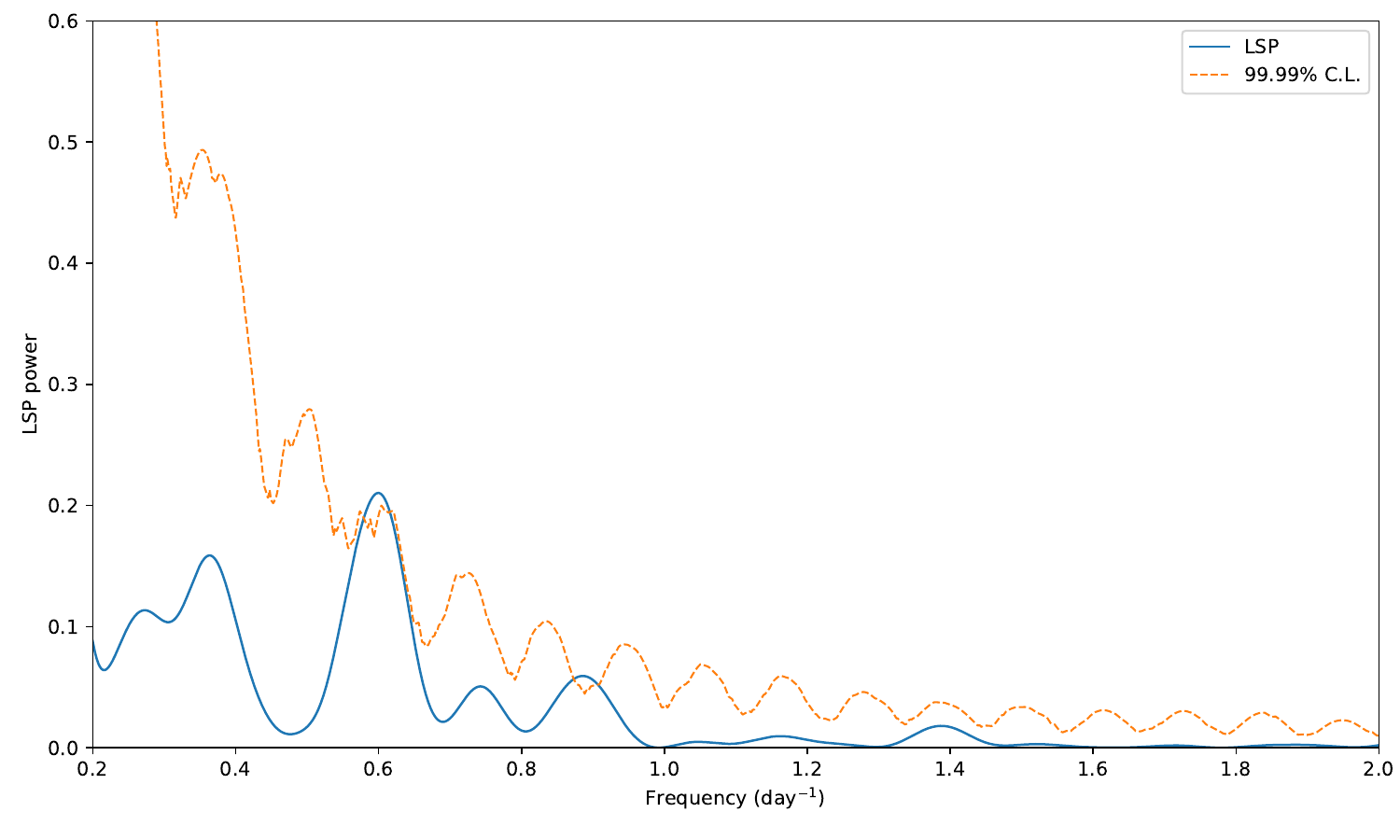}
\caption{Lomb--Scargle periodogram of the TESS light curve of PKS\,0805$-$07. 
A dominant peak is detected at $f=0.6~{\rm d^{-1}}$ ($P\approx1.7$\,d), 
exceeding the 99.99\% confidence level derived from $2\times10^{4}$ Monte Carlo simulations 
following the method of \citet{emmanoulopoulos2013generating}.}
\label{fig:lsp}
\end{figure*}

The resulting periodogram shows a dominant peak at $0.6003^{+0.036}_{-0.032}\,\mathrm{d^{-1}}$, corresponding to a period of $1.67^{+0.09}_{-0.08}\,\mathrm{d}$. (Figure~\ref{fig:lsp}). 
The uncertainty on the detected periodicity is estimated 
using the Monte Carlo simulations described in 
Section~\ref{sig_ev}. For each simulated light curve, the 
LSP is computed over the same frequency range as the 
observed data, and the peak frequency is identified within 
a local window around the observed peak. The distribution 
of these recovered peak frequencies is then used to 
estimate the uncertainty, adopting the 16th and 84th 
percentiles as the $1\sigma$ confidence interval. We note 
that these simulation-based uncertainties reflect the 
dispersion of red-noise peak positions within a frequency 
window around the observed peak, and therefore represent a 
conservative upper bound on the true period uncertainty 
rather than a signal recovery uncertainty. A rigorous 
estimate would require injection-recovery simulations, in 
which a periodic (or quasi-periodic) component with the 
observed amplitude and period is added to each red-noise 
realisation; the dispersion of the recovered periods would 
then provide the actual uncertainty on the periodicity. 
Given the transient and candidate nature of the detected 
modulation, we regard the quoted uncertainties as a 
conservative but reliable estimate of the period 
uncertainty.

Some excess power is also present at lower frequencies
($\sim$0.2$-$0.4\,day$^{-1}$); however, as discussed in
Section~\ref{sig_ev}, Monte Carlo simulations show this to be
consistent with the red-noise character of blazar variability and
no additional peaks reach a comparable significance level. To quantify the statistical significance of the detected periodicity, we computed the false-alarm probability (FAP) using the \texttt{LombScargle.false\_alarm\_probability()} routine from the \texttt{astropy.timeseries} module with \texttt{method="baluev"}. This method provides an analytic estimate of the FAP based on the extreme-value statistics formalism developed by \citet{2008MNRAS.385.1279B}, accounting for the number of independent frequencies sampled by the periodogram. For the dominant peak at $0.6003^{+0.036}_{-0.032}\,\mathrm{d^{-1}}$, we obtain ${\rm FAP} = 1.07\times10^{-4}$, indicating that the probability of the signal arising from stochastic fluctuations is very low and supporting its interpretation as a statistically significant quasi-periodic feature. We note that the \citet{2008MNRAS.385.1279B} method assumes white Gaussian noise
and may underestimate the false alarm probability in the presence
of correlated red-noise variability. This analytic FAP is therefore
reported as an initial diagnostic only; the primary significance
assessment accounting for red-noise is presented in
Section~\ref{sig_ev}.

\begin{figure}
    \centering
    \includegraphics[width=0.5\textwidth]{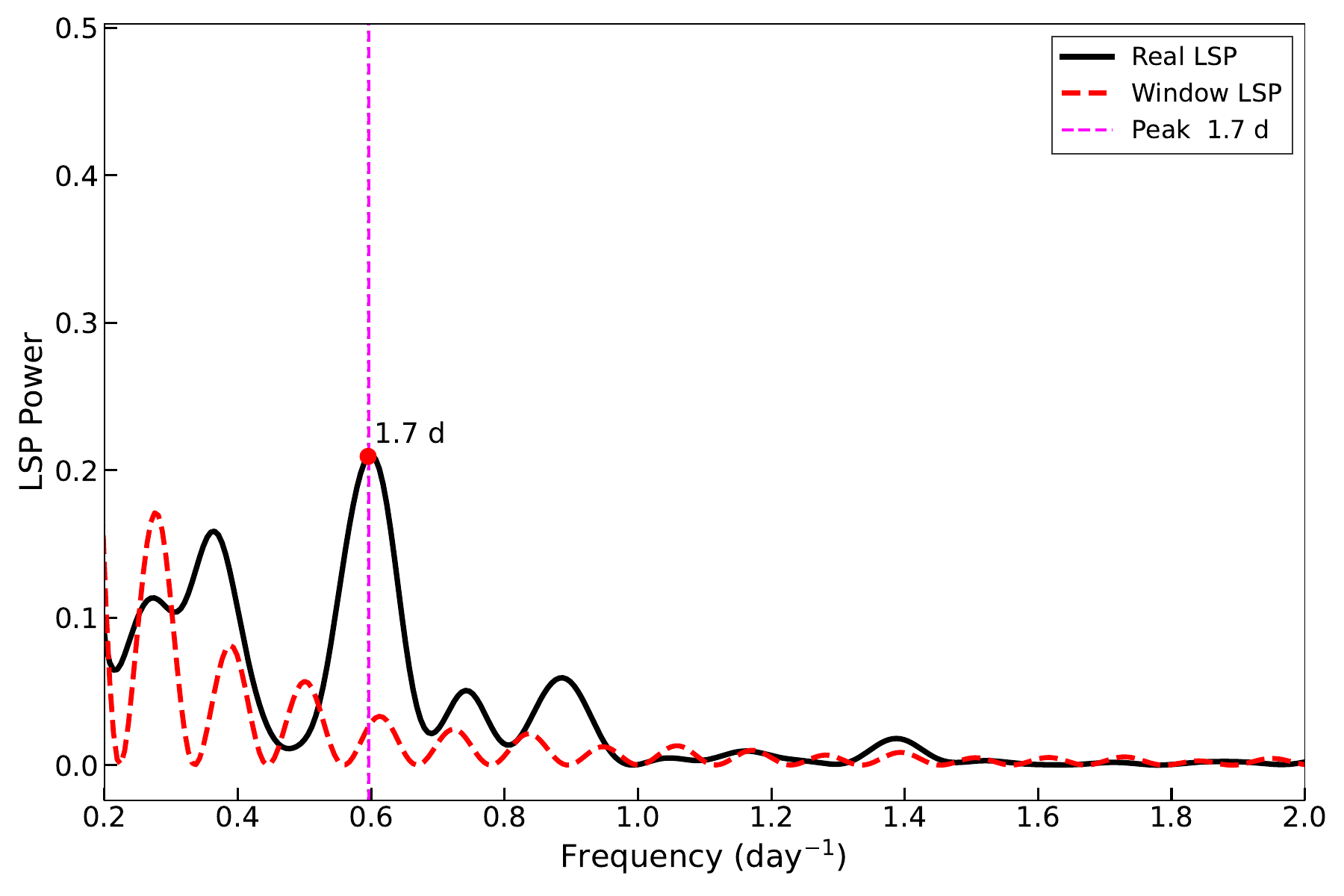}
    \caption{LSP of the TESS light curve.
             The solid black curve shows the real LSP computed from the
             flux measurements. The red dashed curve shows the window
             function LSP, obtained by applying the same algorithm to a
             sequence of unit values at the observed timestamps; it has
             been scaled to the peak power of the real LSP for visual
             comparison. Peaks in the real LSP that coincide with peaks
             in the window function are likely aliases of the sampling
             pattern rather than intrinsic signals. The dominant peak
             at $f \approx 0.6$\,d$^{-1}$ (corresponding to a period
             of $P \approx 1.7$\,d, marked by the red circle) does
             not coincide with any significant window-function peak,
             confirming it as an intrinsic periodic signal in the data.}
    \label{fig:lsp2}
\end{figure}

To distinguish intrinsic periodic
signals from artifacts introduced by the irregular sampling pattern, we also
computed the window function LSP by applying the same algorithm to a sequence
of unit values at the observed timestamps \citep{vanderplas2018understanding}. The window
function LSP was scaled to the peak power of the real LSP for visual
comparison and is shown alongside the real periodogram in
Figure~\ref{fig:lsp2}. Peaks in the real LSP that coincide with peaks in the
window function are candidate aliases of the sampling cadence rather than
astrophysical signals. The dominant peak at $0.6003^{+0.036}_{-0.032}\,\mathrm{d^{-1}}$, corresponding to a period of $1.67^{+0.09}_{-0.08}\,\mathrm{d}$, does not coincide with
any significant window-function peak, confirming that it reflects an
intrinsic periodicity in the light curve.

\begin{figure*}
\centering
\includegraphics[width=0.9\textwidth]{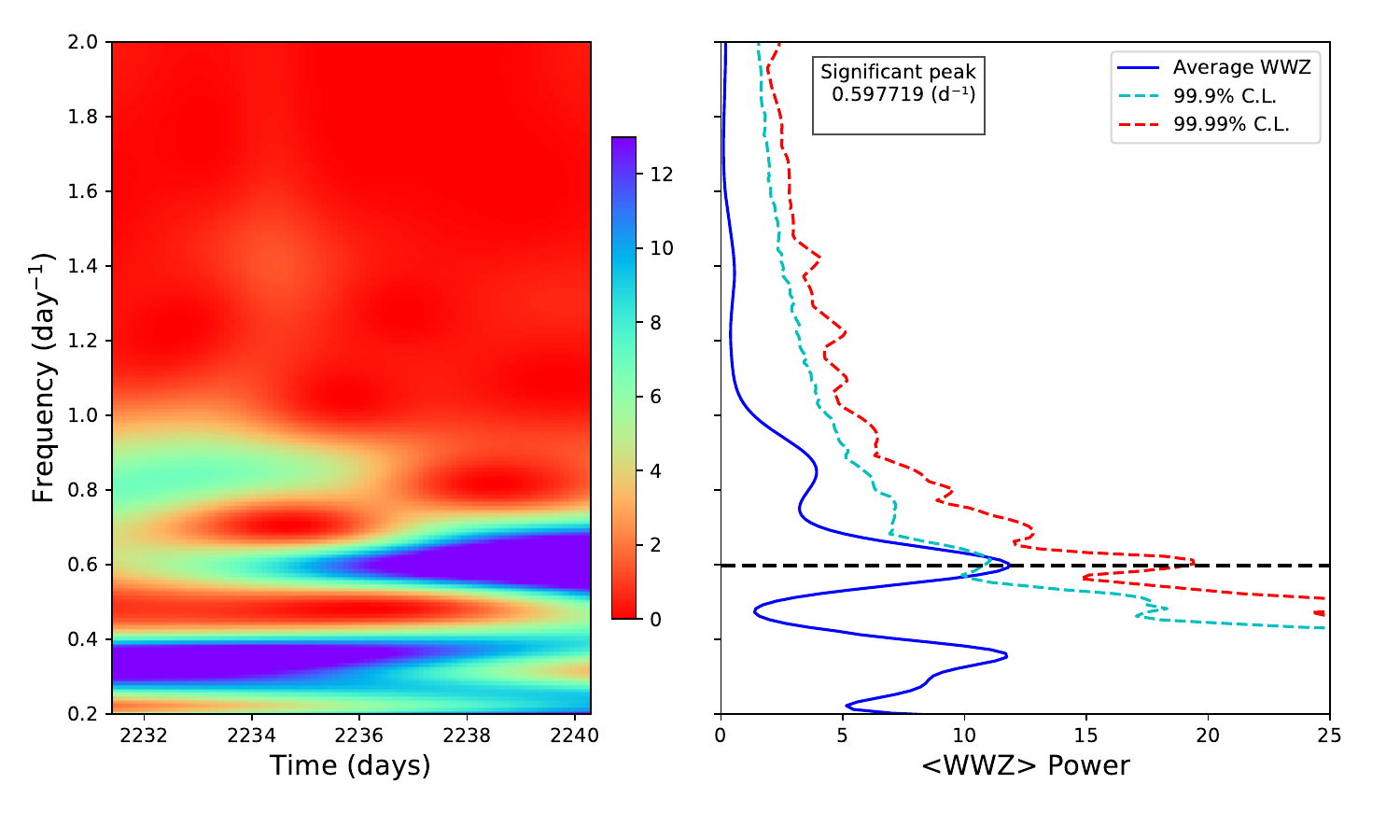}
\caption{Left: WWZ time--frequency map of the TESS light curve showing the evolution of power as a function of time (days) and frequency. 
Right: Average WWZ power spectrum with confidence levels derived from $10^{4}$ Monte Carlo simulations. 
The dashed cyan and red curves denote the 99.9\% and 99.99\% confidence levels, respectively. 
The dashed black line marks the dominant peak at $f=0.5977~{\rm d^{-1}}$ ($P\approx1.67$\,d), 
which exceeds the 99.9\% confidence threshold.}
\label{fig:wwz}
\end{figure*}

\subsection{Weighted Wavelet Z-Transform (WWZ)}\label{sec:wwz}

The weighted wavelet $Z$-transform (WWZ; \citealt{foster1996wavelets}) provides a time--frequency representation of an unevenly sampled light curve by convolving the data with a localized oscillatory kernel. This method is particularly effective for detecting quasi-periodic signals whose strength varies with time, as it simultaneously constrains the characteristic timescale and the temporal interval over which the modulation is present. A genuine periodic component is expected to produce a localized enhancement in WWZ power that evolves as the signal strengthens or weakens.

In this work, we adopted the abbreviated Morlet kernel,
\begin{equation}
f[\omega(t-\tau)] = \exp\!\left[i\omega(t-\tau)-c\,\omega^{2}(t-\tau)^{2}\right],
\end{equation}
and computed the corresponding WWZ projection
\begin{equation}
W[\omega,\tau:x(t)] = \omega^{1/2}\int x(t)\,f^{*}[\omega(t-\tau)]\,dt,
\end{equation}
where $f^{*}$ denotes the complex conjugate of the kernel, $\omega$ is the angular frequency, and $\tau$ represents the time offset. The analysis was performed using the publicly available Python
implementation of the WWZ algorithm.\footnote{\url{https://github.com/eaydin/WWZ}}
The abbreviated Morlet kernel was adopted with a tuning parameter
of $c = 0.001$, which controls the trade-off between time and
frequency resolution \citep{foster1996wavelets}.
The WWZ time--frequency map reveals a localized region of 
enhanced power near $f = 0.5977 \pm 0.020~{\rm d^{-1}}$, 
corresponding to a period of $P \approx 1.67 \pm 0.06$\,d 
(Figure~\ref{fig:wwz}); the frequency uncertainty is 
derived from the simulation-based percentile method. This feature 
is also recovered in the average WWZ spectrum. The absence 
of a continuous ridge of enhanced power across the full 
duration of the light curve indicates that the modulation 
is confined to a limited temporal interval, consistent 
with a transient quasi-periodic signal rather than a 
persistent oscillation. Some power enhancement is also visible at lower frequencies
($\sim$0.2$-$0.4\,day$^{-1}$) in the WWZ map, consistent with
red-noise variability as demonstrated by the Monte Carlo simulations
presented in Section~\ref{sig_ev}.

\begin{figure*}
\centering
\includegraphics[width=0.9\textwidth]{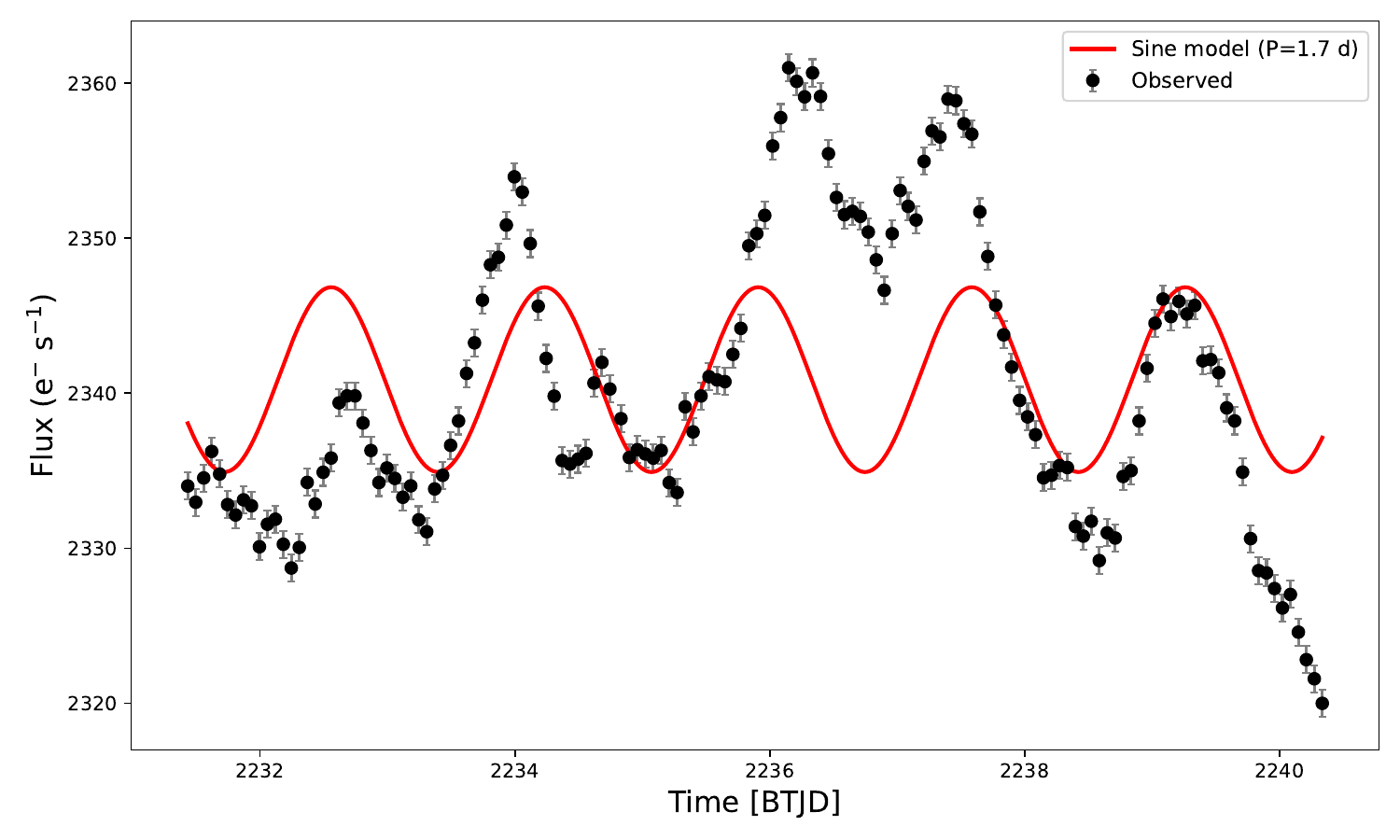}
\caption{Segment of the TESS optical light curve of PKS\,0805$-$07 (1.5\,hr binned) over the interval $\mathrm{BTJD}=2231.4047$--$2240.4047$, showing the quasi-periodic modulation. The black points with error bars represent the observed flux, while the red curve shows the best-fitting sinusoidal model with a period of $P\approx1.7$\,d. The model reproduces the repeating flux enhancements over $\sim5$ cycles, consistent with the timescale identified in the LSP and WWZ analyses.}
\label{fig:sinefit}
\end{figure*}

\begin{figure*}
    \centering
    \includegraphics[width=\textwidth]{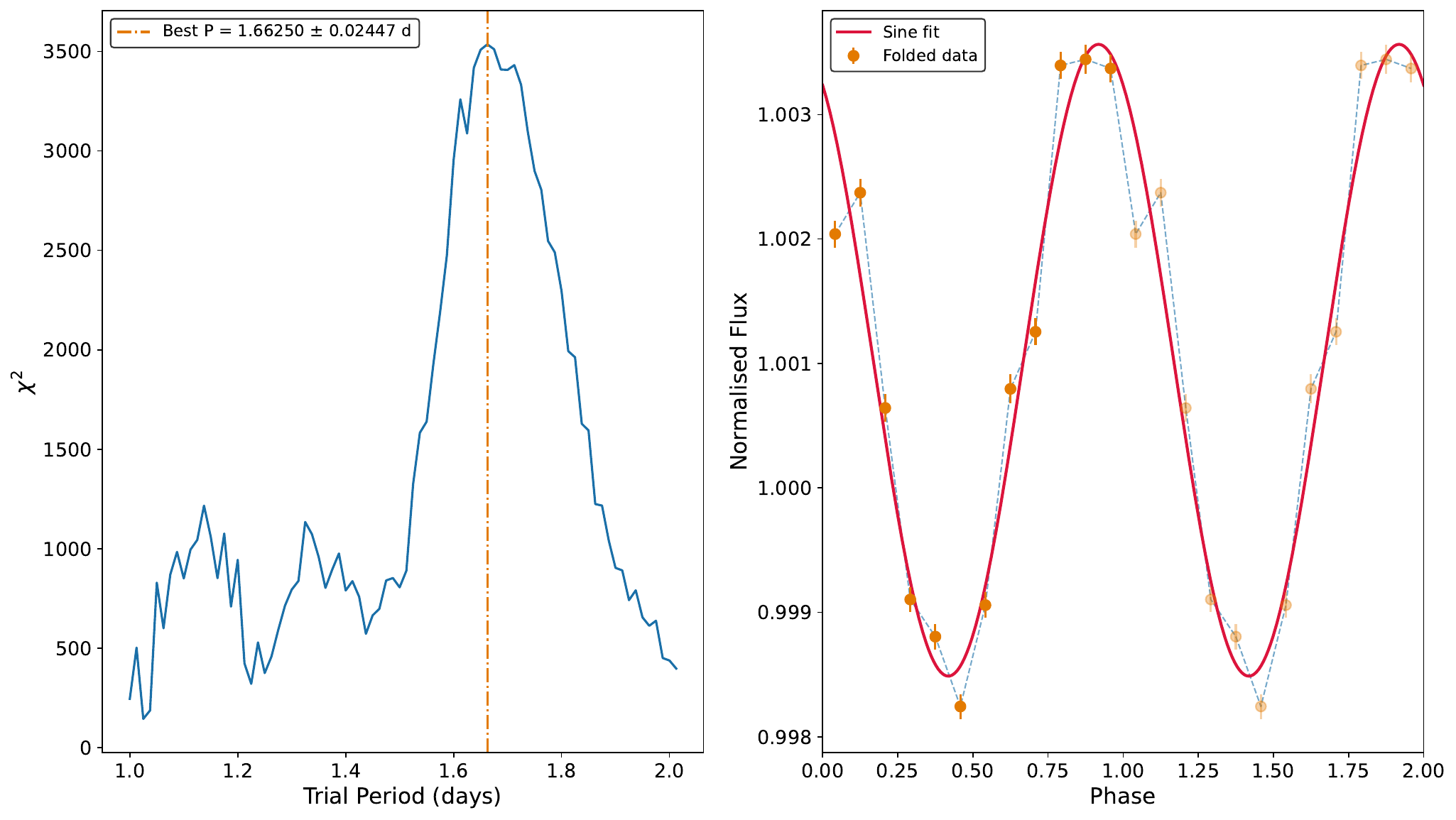}
    \caption{Epoch folding analysis of the TESS light curve. Left panel:
$\chi^2$ statistic as a function of trial period (Equation~\ref{eq:chi2}).
The vertical orange dash-dotted line marks the best period
$P = 1.66250 \pm 0.02447$\,days. Right panel: normalized light curve
folded at the best period and displayed over two cycles for clarity.
The red curve shows the best-fitting sinusoidal model. The faded points
in the second cycle are a repetition of the first; the dashed line
connects the data points. Error bars represent the
standard error of the weighted mean in each phase bin,
$\sigma_i = \left(\sum_{j \in i} w_j\right)^{-1/2}$, where
$w_j = 1/\sigma_j^2$ \citep{Leahy1983,Larsson1996}.}
    \label{fig:ef}
\end{figure*}

\subsection{Epoch Folding Analysis}

To search for periodic signals in the light curve, we employed the epoch
folding technique as formalized by \citet{Leahy1983} and further developed
by \citet{Larsson1996}. Unlike the discrete Fourier transform, epoch folding
does not assume a sinusoidal pulse shape and is well suited for unevenly
sampled data with gaps \citep{Leahy1983, Bhatta2018}. In this method, the
time series of $N$ observations is folded modulo a trial period $P_{\rm
test}$ and the resulting data are co-added into $N_{\rm b}$ equal phase
bins. The folded profile is then tested for uniformity using the Pearson
$\chi^2$ statistic \citep{Leahy1983, Larsson1996}:

\begin{equation}
    \chi^2 = \sum_{i=1}^{N_{\rm b}}
             \frac{\left(x_i - \bar{x}\right)^2}{\sigma_i^2},
    \label{eq:chi2}
\end{equation}

\noindent where $x_i$ is the weighted mean flux in phase bin $i$:

\begin{equation}
    x_i = \frac{\displaystyle\sum_{j \in i} w_j\, f_j}
               {\displaystyle\sum_{j \in i} w_j},
    \qquad w_j = \frac{1}{\sigma_j^2},
    \label{eq:xmean}
\end{equation}

\noindent with the sum running over all observations $j$ falling in bin $i$,
$f_j$ is the flux of observation $j$, and $\sigma_j$ is its $1\sigma$
measurement uncertainty. The quantity $\bar{x}$ is the global weighted mean
across all phase bins:

\begin{equation}
    \bar{x} = \frac{\displaystyle\sum_{i} w_i^{\rm bin}\, x_i}
                   {\displaystyle\sum_{i} w_i^{\rm bin}},
    \qquad w_i^{\rm bin} = \frac{1}{\sigma_i^2}.
    \label{eq:xbar}
\end{equation}

\noindent \citet{Larsson1996} notes that $\sigma_i$ can be computed either
from the variance of data points within each bin, giving $\sigma_i =
\sigma_{\rm tot}/\sqrt{n_i}$ under the null hypothesis, or from individual
measurement uncertainties when these are available. Since the TESS light
curve provides per-observation flux uncertainties $\sigma_j$, we adopt the
latter approach:

\begin{equation}
    \sigma_i = \frac{1}{\sqrt{\displaystyle\sum_{j \in i} w_j}}.
    \label{eq:sigma_i}
\end{equation}

\noindent Under the null hypothesis of no periodic signal, the $\chi^2$
statistic follows a chi-squared distribution with $\nu = N_{\rm b} - 1$
degrees of freedom \citep{Leahy1983}. A value of $\chi^2$ significantly
exceeding $\nu$ signals the presence of a periodic modulation in the light
curve.

The raw TESS flux was normalized by dividing by its median value, yielding
a dimensionless light curve with a mean of unity. This normalization ensures
that the period uncertainty is expressed in physically meaningful units of
time. We searched trial periods in the range $1.0$--$2.0$\,days with a
uniform step size of $\Delta P \approx 0.0125$\,days ($\sim$18\,minutes),
corresponding to one-fifth of the observing cadence. This step size is
sufficiently fine to properly sample the $\chi^2$ peak without missing the
true maximum \citep{Larsson1996}. A total of $N_p = 82$ independent trial
periods were searched and the number of phase bins was set to $N_{\rm b} =
12$, which is a standard choice for epoch folding analyses of unevenly
sampled astronomical light curves \citep{Leahy1983, Bhatta2018}.

The statistical significance of the $\chi^2$ peak was evaluated following
\citet{Leahy1983}. The local significance at the best-period trial
quantifies the probability that the observed $\chi^2$ value arises by chance
from pure noise at that specific period:

\begin{equation}
    \text{Local significance} =
    1 - P\!\left(\chi^2_\nu > \chi^2_{\rm obs}\right),
    \label{eq:local_sig}
\end{equation}

\noindent where $P(\chi^2_\nu > \chi^2_{\rm obs})$ is the survival function
of the $\chi^2$ distribution with $\nu = N_{\rm b} - 1$ degrees of freedom,
evaluated at the observed peak value $\chi^2_{\rm obs}$. Since $N_p = 82$
trial periods were searched rather than a single known period, we also
computed the global significance to account for the look-elsewhere effect
via the Bonferroni correction \citep{Leahy1983}:

\begin{equation}
    \text{Global significance} =
    1 - \left[1 - P\!\left(\chi^2_\nu >
    \chi^2_{\rm obs}\right)\right]^{N_p}.
    \label{eq:global_sig}
\end{equation}

\noindent This gives the probability that no spurious peak as large as
$\chi^2_{\rm obs}$ would arise anywhere across the full set of $N_p$ trials
under the null hypothesis. The corresponding local and global detection
thresholds at the 99\,\% confidence level are obtained from the inverse
$\chi^2$ distribution:

\begin{equation}
    S_{\rm thr}^{\rm local} = F^{-1}_{\chi^2_\nu}(0.99),
    \label{eq:thr_local}
\end{equation}

\begin{equation}
    S_{\rm thr}^{\rm global} =
    F^{-1}_{\chi^2_\nu}\!\left(1 -
    \left[1 - 0.99\right]^{1/N_p}\right),
    \label{eq:thr_global}
\end{equation}

\noindent where $F^{-1}_{\chi^2_\nu}$ denotes the quantile function of the
$\chi^2$ distribution with $\nu$ degrees of freedom.

The $1\sigma$ uncertainty in the best-fit period was estimated using the
analytical procedure of \citet{Larsson1996}. The folded pulse profile at
the best period was mean-subtracted and then decomposed into Fourier
harmonics \citep{Larsson1996}:

\begin{equation}
    x(t) - \bar{x} =
    \sum_{k=1}^{m} A_k
    \cos\!\left(k\,\omega_1\,t - \phi_k\right),
    \label{eq:fourier}
\end{equation}

\noindent where $\omega_1 = 2\pi/P$, $A_k$ and $\phi_k$ are the amplitude
and phase of the $k$-th harmonic respectively, and $m = \lfloor N_{\rm
b}/2 \rfloor - 1$ is the maximum number of independent harmonics. The raw
Fourier amplitudes are attenuated by the finite width of the phase bins.
This bias was corrected using the factor \citep{Larsson1996}:

\begin{equation}
    g(N_{\rm b}) = \frac{1}{\displaystyle\sqrt{
        1 - \frac{\pi^2}{3N_{\rm b}^2}
          + \frac{2\pi^4}{45N_{\rm b}^4}}},
    \label{eq:gNb}
\end{equation}

\noindent by which each raw amplitude is multiplied to recover the true
amplitude $A_k$. The $1\sigma$ period uncertainty is then
\citep{Larsson1996}:

\begin{equation}
    \sigma_P^2 = \frac{6\,\sigma_{\rm tot}^2\,P^4}
        {\pi^2\,N\,T^2\,\displaystyle\sum_{k=1}^{m} k^2 A_k^2},
    \label{eq:sigmaP}
\end{equation}

\noindent where $\sigma_{\rm tot}$ is the standard deviation of the
unfolded normalized light curve, $N$ is the total number of observations,
$T$ is the total time span of the observations, and $P$ is the best-fit
period. Equation~(\ref{eq:sigmaP}) generalizes the single-harmonic
sinusoidal case by making full use of all significant Fourier components of
the pulse shape. For a purely sinusoidal signal it reduces to the
least-squares frequency error \citep{Larsson1996}.

The epoch folding analysis reveals a strong periodic signal at $P =
1.66250 \pm 0.02445$\,d. The peak $\chi^2$
value of $3535.77$ greatly exceeds both the local 99\,\% threshold of
$S_{\rm thr}^{\rm local} = 24.7$ and the global 99\,\% threshold of
$S_{\rm thr}^{\rm global} = 36.8$, for $\nu = 11$ degrees of freedom. Both the local and global significance exceed 99.999\,\%, 
confirming the detection at an extremely high confidence level. The $\chi^2$ periodogram and the folded pulse
profile at the best period are shown in Figure~\ref{fig:ef}. To verify the reliability of this result against red-noise 
variability, we additionally evaluated the significance of 
the epoch-folding peak using $10^4$ Monte Carlo simulations 
following \citet{emmanoulopoulos2013generating}, consistent with the 
treatment adopted for the LSP and WWZ analyses 
(Section~\ref{sig_ev}). Artificial light curves 
reproducing the observed power spectral density and flux 
distribution were generated, and the full $\chi^2$ scan was 
performed on each realization to build a null distribution 
of peak $\chi^2$ values. The observed peak 
$\chi^2 = 3535.77$ exceeds the simulation-derived 99\% 
confidence threshold ($\chi^2_{99\%} = 2880.04$), 
corresponding to a Monte Carlo significance of 99.79\%, 
confirming that the detection significance remains 
consistent across all three independent methods. The folded 
pulse profile is well described by a sinusoidal model, as 
shown in the right panel of Figure~\ref{fig:ef}

\section{Significance Evaluation for LSP and WWZ analysis}
\label{sig_ev}

The variability observed in blazar light curves is typically
dominated by red-noise processes reflecting stochastic fluctuations
arising in the jet or accretion flow. To evaluate the statistical
significance of the candidate periodic feature, we performed Monte
Carlo simulations following the method of
\citet{emmanoulopoulos2013generating}. The PSD of the observed light curve was modelled using a bending
power law, $P(\nu) \propto \nu^{-1}[1 + (\nu/\nu_b)^{\alpha-1}]^{-1}$,
which was preferred over a simple power law by the Bayesian
Information Criterion \citep{2010MNRAS.402..307V}, with best-fit slope
$\alpha = 3.12 \pm 0.06$ and break frequency $\nu_b = 0.567 \pm
0.019$\,day$^{-1}$, consistent with the red-noise character of
blazar variability. The PDF was estimated directly from the flux
distribution of the observed data. Artificial light curves
reproducing both the best-fit PSD and the observed PDF were
generated and sampled identically to the TESS data.

For the LSP analysis, $2\times10^{4}$ simulated light curves were
produced, and the distribution of periodogram powers at each trial
frequency was used to derive confidence levels. The dominant peak
at $f\approx0.6~\mathrm{d^{-1}}$ ($P\approx1.7$\,d) exceeds the
99.99\% confidence threshold (Figure~\ref{fig:lsp}). For the WWZ
analysis, $10^{4}$ simulated light curves were generated, and the
average WWZ spectrum was used to estimate the significance of the
detected features. A consistent peak is recovered at
$f\approx0.5977~\mathrm{d^{-1}}$ ($P\approx1.67$\,d), reaching
the $\sim99.9$\% confidence level (Figure~\ref{fig:wwz}). The
agreement between the LSP and WWZ results, together with their
consistency with the red-noise simulations, supports the
interpretation of the detected modulation as a statistically
significant, albeit transient, quasi-periodic signal. The detected
period corresponds to approximately five complete cycles in the
light curve (see Figure~\ref{fig:sinefit}), which further supports
its quasi-periodic nature. Early studies demonstrated that apparent
periodic patterns spanning only a few cycles can arise from
stochastic (flicker-noise) processes \citep{1978ComAp...7..103P}.
Subsequent work showed that the statistical reliability of a
periodic detection increases with the number of coherent cycles,
with $\gtrsim5$ cycles providing substantially stronger evidence
against a purely stochastic origin, whereas signals with only
$\sim2$ cycles are generally indistinguishable from red-noise
variability \citep{2016MNRAS.461.3145V}. In this context, the
$\sim1.7$\,d modulation detected in PKS\,0805$-$07 spans
approximately five cycles (Figure~\ref{fig:sinefit}), supporting
its interpretation as a quasi-periodic feature rather than a
stochastic fluctuation. Nevertheless, we caution that the observed
peak only marginally exceeds the highest confidence level in the
WWZ analysis, and that red-noise simulations can occasionally
produce comparable power at similar frequencies. The detected
feature should therefore be regarded as a \textit{candidate}
quasi-periodic oscillation, and additional high-cadence
observations are required to confirm its recurrence.

\section{Summary and Discussion}
\label{sec:sum}

We performed a timing analysis of the high-cadence optical 
light curve of PKS\,0805$-$07 obtained with TESS during 
Sector~34 (MJD~59229--59254; 10-min cadence). The 
systematics-corrected light curve was extracted using the 
\textsc{Quaver} pipeline, from which both the FHO and SPO 
reductions were generated; the PCA-corrected light curve was 
adopted for the timing analysis. Owing to the large number of 
data points, the light curve was rebinned by averaging 
measurements within 1.5\,hr intervals for ease of subsequent 
time-series processing. The period search was carried out over 
the interval BTJD $=2231.4047$--$2240.4047$ 
(MJD $=59230.90$--$59239.90$). The Lomb--Scargle periodogram 
reveals a dominant peak at 
$0.6003^{+0.036}_{-0.032}\,\mathrm{d^{-1}}$, corresponding 
to a period of $1.67^{+0.09}_{-0.08}\,\mathrm{d}$, with a 
false-alarm probability of $1.07\times10^{-4}$ (Baluev 
method) and exceeding the $99.99\%$ confidence level derived 
from Emmanoulopoulos simulations. The weighted wavelet 
$Z$-transform independently recovers a consistent timescale 
at $f\approx0.5977~\mathrm{d^{-1}}$ ($P\approx1.67$\,d) 
with $\sim99.9\%$ significance and shows that the modulation 
is localized in time rather than persistent across the full 
light curve. Epoch folding independently confirms a periodic 
signal at $P = 1.66250 \pm 0.02447$\,d, with a peak 
$\chi^2 = 3535.77$ exceeding both the local and global 99\% 
analytical thresholds and the simulation-derived 99\% 
confidence threshold ($\chi^2_{99\%} = 2880.04$) from 
$10^4$ Emmanoulopoulos Monte Carlo simulations, yielding a 
Monte Carlo significance of 99.79\%. A sinusoidal model 
fitted to the 1.5\,hr binned data reproduces the repeating 
flux enhancements over $\sim5$ coherent cycles, indicating a 
transient quasi-periodic feature superposed on the underlying 
stochastic variability characteristic of blazars. Given that optical blazar power spectra are typically dominated by red-noise processes, the detection of a coherent day-scale modulation in uniformly sampled TESS data provides strong statistical evidence for a quasi-periodic feature.

Short-timescale QPOs in blazars are difficult to interpret within models that are commonly invoked for year-scale periodicities, such as binary supermassive black hole systems \citep{2008Natur.452..851V,2021JApA...42...92L}, long-term jet precession \citep{2004ApJ...615L...5R,2018MNRAS.474L..81L}, or Lense--Thirring precession of the accretion flow \citep{1998ApJ...492L..59S}, because those mechanisms naturally produce much longer characteristic periods. The $\sim$1.7\,day timescale therefore points to processes operating in compact regions of the accretion flow or the inner relativistic jet. Below we discuss two physically plausible scenarios.

\subsection*{a) Disk-based hotspot or inner accretion-flow oscillations}

One plausible interpretation for intraday quasi-periodic variability is orbital motion of a localized hotspot or other non-axisymmetric structure near the innermost stable circular orbit (ISCO) of the accretion disk \citep{1993ApJ...411..602C,1993ApJ...406..420M}. In this picture, transient inhomogeneities such as spiral shocks or pulsation modes can modulate the optical emission on dynamical timescales \citep{2008ApJ...679..182E,2023MNRAS.520.4118C}. Assuming that the observed period corresponds to the orbital timescale at the ISCO, the black hole mass can be estimated using the standard relation \citep{2009ApJ...690..216G}.

\[
\frac{M_{\mathrm{BH}}}{M_{\odot}}=\frac{3.23\times10^{4}\,P}{\left(r^{3/2}+a\right)(1+z)} ,
\]
where $P$ is the period in seconds, $z=1.837$ is the redshift of PKS\,0805$-$07 \citep{1988ApJ...327..561W}, $r$ is the ISCO radius in units of $GM/c^{2}$, and $a$ is the dimensionless spin parameter. Adopting the ISCO orbital interpretation, the black hole mass can be estimated using the standard relation. For a Schwarzschild black hole ($r=6$, $a=0$) we obtain $M_{\rm BH}\approx1.1\times10^{8}\,M_{\odot}$, while for a maximally rotating Kerr black hole ($r=1.2$, $a=0.9982$) the inferred mass is $M_{\rm BH}\approx7.2\times10^{8}\,M_{\odot}$ \citep{2009ApJ...690..216G,2008ApJ...679..182E}. These values lie within the typical mass range reported for FSRQs. This scenario naturally explains the day-scale modulation as a transient disk phenomenon; however, in jet-dominated blazars the optical emission is often primarily synchrotron radiation from the relativistic jet, which may reduce the contribution of disk-based variability.

\subsection*{b) Jet-based kink instability}

An alternative and physically compelling explanation is that the quasi-periodic signal originates within the relativistic jet through the development of magnetohydrodynamic kink instabilities \citep{2020MNRAS.494.1817D,2009ApJ...700..684M}. In a jet permeated by a helical or toroidal magnetic field, current-driven kink modes can produce transverse displacements of the plasma and distort the magnetic field geometry \citep{2009ApJ...700..684M}. The associated dissipation of magnetic energy leads to enhanced particle acceleration and localized increases in synchrotron emission \citep{2020MNRAS.494.1817D}. As the kink grows and propagates, quasi-periodic compressions of the emitting region can generate flux modulations on characteristic timescales related to the kink growth time \citep{2020MNRAS.494.1817D,2022Natur.609..265J}. Such behavior has been linked to rapid quasi-periodic variability observed in relativistic jets and has been explored in recent optical TESS blazar studies \citep{2022Natur.609..265J,2024MNRAS.527.9132T}.
The growth rate of a kinked structure in a relativistic jet can be estimated by quantifying its transverse displacement and propagation speed \citep{2009ApJ...700..684M}. In the kink-instability framework, the characteristic growth time $\tau_{\rm KI}$ is given by the ratio of the lateral displacement of the emitting region from the jet axis, $R_{\rm KI}$, to the mean transverse propagation velocity $\langle v_{\rm tr}\rangle$ \citep{2020MNRAS.494.1817D}. This growth timescale is expected to be comparable to the observed quasi-periodic modulation.

In the observer’s frame, the characteristic period associated with the kink instability is given by \citep{2020MNRAS.494.1817D}
\begin{equation}
T_{\rm obs} = \frac{R_{\rm KI}}{\langle v_{\rm tr}\rangle\,\delta},
\end{equation}
where $\delta$ is the Doppler factor and $R_{\rm KI}$ represents the size of the emitting region in the co-moving frame. The resulting timescale therefore depends on key jet parameters, including the viewing angle (through $\delta$), the bulk flow speed, and the characteristic size of the emission region. Consequently, the observed quasi-periodicity reflects the dynamical evolution of magnetized plasma structures within the relativistic jet rather than a purely geometric modulation.
Adopting representative blazar parameters for the kink-instability scenario, with a Doppler factor $\delta = 15$, a transverse propagation speed $\langle v_{\rm tr}\rangle\approx0.16c$ \citep{2020MNRAS.494.1817D}, and an emitting-region size $R_{\rm KI} = 10^{16}$--$10^{17}$\,cm, the expected observer-frame timescale lies in the range $\sim1.6$--$16$\,days. The observed optical modulation at $P\approx1.7$\,days is located at the lower end of this interval, indicating that the quasi-periodic signal is consistent with a compact emitting region of size $R_{\rm KI}\sim10^{16}$\,cm. This supports a scenario in which localized kink-induced compressions within the relativistic jet enhance the synchrotron emission on characteristic dynamical timescales. This consistency supports a jet-based origin in which periodic plasma compression associated with the growth of a kink produces the observed short-timescale variability.
Kink instabilities are not expected to operate as a strictly steady process in relativistic jets, since their development depends on the
time-dependent injection of magnetic energy and plasma into the flow. As a result, kink-driven modulations may persist only for a limited
number of cycles rather than producing a long-lived periodic signal. This behavior is consistent with the WWZ results, which show that the
$\sim1.7$\,d modulation is localized in time and does not extend across the entire light curve.

Other disk-based mechanisms could, in principle, produce short-timescale quasi-periodic variability. Normal modes of oscillations trapped in the inner accretion flow by strong gravity \citep[e.g.,][]{1997ApJ...476..589P,2008ApJ...679..182E} and turbulence driven by magnetorotational instability (MRI) \citep[e.g.,][]{2004ApJ...609L..63A} may generate transient modulations in the emitted flux. Lense--Thirring precession of a tilted inner disk can also introduce quasi-periodic signals \citep{1998ApJ...492L..59S}, although it typically operates on longer (weeks to months) timescales. However, these disk-dominated processes are expected to be more relevant in Seyfert galaxies and radio-quiet quasars, where the optical emission originates primarily from the accretion flow. In blazars such as PKS\,0805$-$07, the optical band is strongly jet dominated, making a disk origin less favorable for the observed day-scale modulation.

\subsection*{c) Compact SMBHB and TDE Scenarios}

\citet{XinHaiman2021} showed that orbital periods of 1--2\,days 
correspond to the most compact SMBHBs expected in large quasar 
surveys such as the Legacy Survey of Space and Time (LSST), 
which will enter the Laser Interferometer Space Antenna (LISA) 
mHz gravitational wave band within 5--15\,years --- the 
so-called ``LISA verification binaries,'' with typical total 
masses of $\sim10^{7}$--$10^{8}\,M_{\odot}$. In this picture, 
the quasi-sinusoidal modulation would arise from relativistic 
Doppler boosting of minidisk emission \citep{DOrazio2015}, with 
the period reflecting the binary orbital period. More recently, 
\citet{Xin2026} developed a fully Bayesian framework to detect 
the characteristic electromagnetic ``chirp'' --- a progressive 
increase in orbital frequency driven by gravitational wave 
emission --- in LSST quasar lightcurves, demonstrating that 
such chirps are in principle detectable for binaries with times 
to merger as long as $10^{4}$\,yr. However, several arguments 
disfavour this interpretation for the present detection. First, 
such systems are intrinsically rare: \citet{XinHaiman2021} 
predict only $\sim$10--150 LISA-detectable compact binaries 
among $\sim$100 million LSST quasars, making the 
\textit{a priori} probability of a single-source TESS detection 
extremely low. Second, the black hole mass inferred for 
PKS\,0805$-$07 --- $M_{\rm BH}\sim1.1\times10^{8}\,M_{\odot}$ 
(Schwarzschild) to $7.2\times10^{8}\,M_{\odot}$ (maximally 
spinning Kerr) --- places the source at or above the upper 
boundary of the LISA verification binary mass range of 
\citet{XinHaiman2021}, making the compact SMBHB interpretation 
quantitatively strained, particularly for the Kerr case. Third, 
a gravitational-wave-driven binary produces a persistent, 
continuously chirping signal with $\dot{f}>0$ \citep{Xin2026}, 
fundamentally inconsistent with the transient modulation 
confined to a single $\sim$9\,day window and absent in all 
other available TESS sectors (7, 61, 88, and 99). Fourth, the 
optical emission of PKS\,0805$-$07 is strongly jet-dominated, 
making it unlikely that Doppler-boost modulation from a binary 
would produce a detectable fractional variation against the 
dominant synchrotron component. Fifth, the $\sim$255 and 
$\sim$112\,day $\gamma$-ray quasi-periodicities previously 
reported for this source \citep{zxgv-fzv5} are most naturally 
attributed to geometric effects in the jet rather than a binary 
orbital period, since SMBHB-induced QPOs are generally expected 
on timescales of $\sim$1--25\,years \citep{Komossa2006}. Taken 
together, while we cannot formally exclude the compact SMBHB 
scenario on the basis of the present data alone, the transient 
nature of the signal, the jet-dominated optical emission, the 
quantitative tension with the LISA verification binary mass 
range, and the \textit{a priori} rarity of such systems all 
argue strongly against it. A definitive test would require 
multi-sector monitoring and, if recurrence is confirmed, 
measurement of $\dot{f}$ to test the gravitational wave chirp 
relation \citep{Xin2026}.

A tidal disruption event (TDE) origin is similarly disfavoured. 
TDEs and quasi-periodic eruptions (QPEs) strongly prefer 
quiescent or recently faded AGN environments 
\citep{MilanVeres2026,Jiang2025}, in stark contrast to the 
luminous, jet-dominated nature of PKS\,0805$-$07. The detected 
modulation --- a low-amplitude ($\sim$1\%), quasi-sinusoidal 
oscillation over $\sim$5 cycles --- is morphologically 
inconsistent with the smooth TDE rise-and-decline profile 
\citep{MilanVeres2026}; inspection of the full Sector~34 
lightcurve (Figure~\ref{fig:sector34_lc}) reveals no such 
secular brightening superimposed on the stochastic blazar 
continuum. Confirmed TDE-associated QPOs are observed 
exclusively in the X-ray band \citep{Smith2021,Suzuguchi2026}, 
arising from extreme mass-ratio inspiral (EMRI)--disc 
interactions at temperatures of $\sim$1--100\,keV, a mechanism 
entirely distinct from the optical synchrotron variability 
reported here. Furthermore, all confirmed QPE host galaxies 
have $M_{\rm BH}\sim10^{5}$--$10^{7.3}\,M_{\odot}$ 
\citep{Jiang2025}, well below the mass inferred for 
PKS\,0805$-$07; for $M_{\rm BH}\gtrsim10^{8}\,M_{\odot}$ the 
tidal disruption radius falls within the event horizon for 
solar-type stars \citep{Webb2023}, further suppressing a TDE 
origin. We note that the dimensionless frequency product 
$Mf\times GM_{\odot}/c^{3}\approx0.07$ --- computed using 
$M_{\rm BH}\sim7.2\times10^{8}\,M_{\odot}$ and the rest-frame 
QPO frequency $f_{\rm rest}=f_{\rm obs}\times(1+z)\approx 
1.96\times10^{-5}$\,Hz --- is broadly consistent with the TDE 
locus of \citet{Smith2021}, but this coincidence in the 
mass--frequency parameter space does not establish a physical 
identification with the TDE class, since the same $Mf$ range 
is expected from disc instabilities, diskoseismic modes, and 
jet-related processes \citep{Smith2021,Wagoner2001}. We flag 
the absence of simultaneous X-ray monitoring at the Sector~34 
epoch (MJD~59229--59254) as a genuine limitation of the present 
work. Future simultaneous X-ray and UV monitoring 
contemporaneous with any recurrence of the optical QPO would 
provide the most decisive discrimination, since TDE-associated 
QPOs are most prominently detected in the X-ray band 
\citep{Smith2021,Suzuguchi2026}, while jet-driven modulations 
are expected to produce correlated variability across the 
optical, UV, and radio bands \citep{2022Natur.609..265J,
2024MNRAS.527.9132T,2017Natur.552..374R,Tripathib}.

\subsection*{Implications}

The episodic character of the modulation and its occurrence 
over only a limited number of cycles indicate that the 
underlying process is not strictly periodic but rather 
reflects a temporary coherent structure embedded within 
stochastic jet variability. Such transient QPO-like features 
have been reported in other blazars observed with TESS, where 
short-timescale oscillations appear during active phases and 
do not persist across the entire light curve 
\citep{2022Natur.609..265J, 2024MNRAS.527.9132T}. While the presence of 
approximately five coherent cycles provides initial 
evidence against a purely stochastic origin --- since 
red-noise processes rarely sustain quasi-sinusoidal patterns 
over multiple consecutive cycles \citep{2016MNRAS.461.3145V} --- we 
caution that this alone is insufficient for a firm or robust 
confirmation of a genuine periodic signal. As noted by 
\citet{1978ComAp...7..103P}, apparent periodic patterns spanning only 
a few cycles can arise from stochastic flicker-noise 
processes, and \citet{2016MNRAS.461.3145V} demonstrated that 
statistical reliability increases substantially only with 
a larger number of coherent cycles. The present detection 
should therefore be regarded as a \textit{candidate} 
quasi-periodic oscillation rather than a confirmed 
periodicity; additional high-cadence observations are 
essential to establish whether the signal recurs and to 
determine whether it represents a genuine quasi-periodic 
phenomenon or an unusually coherent noise fluctuation.
The duration of a single TESS sector ($\sim$25--27\,d) 
imposes an intrinsic constraint on the range of detectable 
periods. Quasi-periods longer than $\sim$5--6\,d would 
correspond to fewer than four cycles and therefore cannot 
be reliably established \citep{2016MNRAS.461.3145V}, while periods 
shorter than a few hours would approach the Nyquist limit 
of the sampling cadence and lie in the white-noise regime 
of the power spectrum. The detected $\sim$1.7\,d modulation 
falls well within the optimal sensitivity window of the 
data, allowing multiple coherent cycles to be observed. 
Nevertheless, the marginal excess above the highest 
confidence threshold in the WWZ analysis means that the 
red-noise origin cannot be fully excluded on the basis 
of the present data alone, and the significance should 
not be over-interpreted.
The jet-based kink-instability scenario provides a natural 
explanation for both the observed timescale and the 
transient character of the signal \citep{2020MNRAS.494.1817D, 
2022Natur.609..265J, 2024MNRAS.527.9132T}, while the disk-hotspot 
interpretation yields a black hole mass consistent with 
independent FSRQ expectations \citep{2009ApJ...690..216G, 
2008ApJ...679..182E} and therefore cannot be ruled out. 
Distinguishing between these scenarios, and discriminating 
them from a red-noise origin, will require continued 
high-cadence, multiwavelength monitoring. Such observations 
are essential not only to confirm or refute the recurrence 
of the $\sim$1.7\,d oscillation, but also to investigate 
its possible connection with the longer-timescale 
$\gamma$-ray quasi-periodicities previously reported for 
PKS\,0805$-$07 \citep{zxgv-fzv5}, and to establish whether 
short- and long-term quasi-periodicities arise from a 
common physical origin. In particular, if the modulation 
recurs in future TESS sectors with a consistent period, 
that would substantially strengthen the case for a 
quasi-periodic rather than stochastic origin and would 
provide the discriminating power needed to favour one 
physical mechanism over another.

\section{Acknowledgements}
ZS is supported by the Department of Science and Technology, Govt. of India, under the INSPIRE Faculty grant (DST/INSPIRE/04/2020/002319). SA, ZS and  NI express  gratitude to the Inter-University Centre for Astronomy and Astrophysics (IUCAA) in Pune, India, for the support and facilities provided.

\bibliographystyle{elsarticle-harv} 
\bibliography{sample631}





 

\end{document}